\documentclass[runningheads]{llncs}
\usepackage[T1]{fontenc}
\usepackage{mdframed}
\usepackage{subcaption}
\usepackage{graphicx}
\usepackage{comment}
\usepackage{todonotes}
\usepackage{hyperref}
\usepackage{xspace}
\newcommand{\ea}{\emph{et al.}\xspace}
\definecolor{airforceblue}{rgb}{0.36, 0.54, 0.66}
\definecolor{caribbeangreen}{rgb}{0.0, 0.8, 0.6}
\definecolor{coolblack}{rgb}{0.0, 0.18, 0.39}
\newcommand{\MethDescr}[1]{\textcolor{coolblack}{\textit{\textsc{#1}}}}

\usepackage[most]{tcolorbox}

\newtcolorbox{mytextbox}[1][]{%
  sharp corners,
  enhanced,
  colback=white,
  attach title to upper,
  #1
}

\begin{document}
\title{Combining Different Existing Methods for Describing Steganography Hiding Methods}
\titlerunning{Describing Steganography Methods}

\author{Steffen Wendzel\inst{1}\orcidID{0000-0002-1913-5912},
Christian Krätzer\inst{2}\orcidID{0000-0002-0138-4638},
Jana Dittmann\inst{2}\orcidID{0009-0003-7985-8041},
Luca Caviglione\inst{3}\orcidID{0000-0001-6466-3354},
Aleksandra Mileva\inst{4}\orcidID{0000-0003-0706-6355},
Tobias Schmidbauer\inst{5}\orcidID{0000-0001-5912-0857},
Claus Vielhauer\inst{6}\orcidID{0009-0007-7125-2722},
Sebastian Zander\inst{7}\orcidID{0000-0002-2084-7204}
}

\authorrunning{S. Wendzel et al.}

\institute{Ulm University, Germany\\
\email{steffen.wendzel@uni-ulm.com} \and
Otto-von-Guericke University Magdeburg, Germany\\
\email{\{dittmann,kraetzer\}@ovgu.de} \and
Institute for Applied Mathematics and Information Technology, Italy\\
\email{luca.caviglione@ge.imati.cnr.it} \and
Goce Delcev University, N. Macedonia\\
\email{aleksandra.mileva@ugd.edu.mk} \and
Technische Hochschule Nürnberg Georg Simon Ohm\\
\email{tobias.schmidbauer@th-nuernberg.de} \and
Brandenburg University of Applied Sciences, Brandenburg, Germany\\
\email{claus.vielhauer@th-brandenburg.de} \and
Murdoch University, Perth, Australia\\
\email{s.zander@murdoch.edu.au}\\
}%

\maketitle              
%
%
%

\begin{abstract}
The proliferation of digital carriers that can be exploited to conceal arbitrary data has greatly increased the number of techniques for implementing network steganography. As a result, the literature overlaps greatly in terms of concepts and terminology. Moreover, from a cybersecurity viewpoint, the same hiding mechanism may be perceived differently, making harder the development of a unique defensive strategy or the definition of practices to mitigate risks arising from the use of steganography. To mitigate these drawbacks, several researchers introduced approaches that aid in the unified description of steganography methods and network covert channels.

Understanding and combining all descriptive methods for steganography techniques is a challenging but important task. For instance, researchers might want to explain how malware applies a certain steganography technique or categorize a novel hiding approach.
Consequently, this paper aims to provide an introduction to the concept of \textit{descriptive methods for steganography}. The paper is organized in the form of a tutorial, with the main goal of explaining how existing descriptions and taxonomy objects can be combined to achieve a detailed categorization and description of hiding methods. To show how this can effectively help the research community, the paper also contains various real-world examples.

\end{abstract}

\begin{mytextbox}[colupper=blue,fontupper=\bfseries\normalsize]
This is a pre-print. The final version of this paper will be published in the proceedings of the \emph{ARES 2025 Workshops} (CUING Workshop), Springer LNCS.
\end{mytextbox}

\section{Introduction}

The increasing diffusion of digital objects increases the opportunities for hiding data. For instance, the need for enforcing copyright, tracking the diffusion of information, or preventing that data is ingested without a suitable consent, culminated in a vast array of \textit{watermarking} mechanisms~\cite{wan2022comprehensive}. At the same time, the massive softwarization of services in combination with the complexity of modern software supply chains required the design of new watermarks for concealing control information. As an example, software artifacts should be traceable to guarantee provenance and early detect tampering that may cause outages or data breaches~\cite{dey2019software}. In parallel, the ubiquitous diffusion of AI opened up new challenges. Information needs to be hidden in datasets and models to support licensing schemes, protect large monetary investments, and track quality in AI-as-a-Service frameworks~\cite{regazzoni2021protecting}. 

However, the process of hiding information is not always performed for legitimate purposes. This is the case with malware endowed with steganographic capabilities. In essence, this class of malicious software tries to prevent detection or bypass security countermeasures by cloaking configuration data, offensive attack routines, or optional payloads within multimedia assets. As a result, malware is no longer monolithic, but is implemented through a multi-stage architecture, which reduces its footprint~\cite{caviglione2022never}. Another offensive approach concerns the creation of \textit{covert channels}, i.e., parasitic communication paths that can be used to exchange data in a secret and unauthorized manner. Even with real-world attacks abounding in \textit{carriers} that can be abused for the covert communication (e.g., patterns of syscalls or hardware behaviors), the most effective approaches take advantage of network traffic~\cite{Strachanski:Stegomalware}.

As a result, the works dealing with information hiding and steganography largely overlap both in terms of terminology and concepts. In some cases, the same idea is reinvented multiple times, wasting resources and complicating the retrieval of knowledge. 
Another major issue is rooted within the double-edged nature of mechanisms devoted to cloak data. On one hand, they have proven their effectiveness for tracking and copyright purposes (e.g., watermarks). On the other hand, they are definitely becoming an important resource in the toolbox of attackers (e.g., to implement stegomalware). This causes a ``mismatch'' in the perception of the various steganographic approaches, as some ideas may require to be designed contextually with a proper defensive strategy. Moreover, different backgrounds and use cases hinder the possibility of developing general countermeasures. For instance, the mitigation of network covert channels requires preventing ambiguities that could be exploited by an attacker early on, i.e., during the design stage. Unfortunately, this conflicts with the need for offering techniques to mark packet flows and control their route through the Internet, such as for traffic engineering goals~\cite{caviglione2024you}. 

To cope with the aforementioned pitfalls, several authors introduced different approaches for a \textit{unified description} of steganography methods and network covert channels. Specifically, this paper explains how existing steganography taxonomy and description objects can be used jointly to achieve a unified and clear explanation. The goal here is to provide a solid foundation for the scientific literature on steganography and covert channels. 
Our paper is accompanied by an inter-active online tool: \url{https://patterns.omi.uni-ulm.de/desrcovert/}.

In summary, the main contribution of this work is to provide an introduction to concepts related to \textit{descriptive methods for steganography}. The paper provides a tutorial on how existing descriptions and taxonomy objects can be combined for designing precise descriptions of several hiding methods.

The remainder of this paper is structured as follows. We discuss the existing concepts in Section~\ref{sect:basics} and then describe how these can be combined in Section~\ref{sect:meta}. We provide tutorial examples in Section~\ref{sect:examples} and a brief discussion in Section~\ref{sect:disc}. Finally, Section~\ref{sect:concl} concludes.

\section{Related Work and Fundamentals}\label{sect:basics}

This section reviews previous approaches on how steganographic methods and techniques for the creation of covert channels can be organized or described in terms of hiding patterns. For the sake of brevity, this section does not embrace the literature on information hiding on a \textit{tout court} manner. Rather, it should be considered as a complement of more comprehensive works, see, e.g.,~\cite{carrara2016out} for a general overview of cloaked/abusive communication paths. 

\paragraph{Hiding Patterns.}
For the specific case of taxonomies, several authors already categorized information hiding and steganography methods. As a result, a relevant corpus of works has emerged~\cite{bennett2004linguistic,dhawan2021analysis,Fridich2009,johnson2000survey,knochel2024text,li2011survey,lubacz2014principles,majeed2021review,mileva2014covert,petitcolas1999information,provos2003hide,zander2010performance,zander,EntropyTaxonomyNCC}. However, the existing categorizations have either focused on high-level aspects or have been tailored to specific domains. For instance, many works only consider network steganography or image steganography.

In 2015, the concept of \emph{hiding patterns} has been introduced to provide a generic taxonomy on the hiding process \cite{CSUR15}. In essence, hiding patterns describe hiding methods in an abstract fashion. Until recently, the description of hiding patterns still has been domain-specific: the original taxonomy was tailored for network steganography \cite{CSUR15} but an analysis of the taxonomy in the context of cyber-physical systems exists as well \cite{hildebrandt2020information}.

More recently, a \emph{generic pattern-based taxonomy for all steganography domains} has been introduced \cite{CSUR25}. Based on this taxonomy, hiding patterns either describe how a secret message is \emph{embedded} in a carrier (so-called \emph{embedding patterns}) or how the secret message is \emph{represented} (so-called \emph{representation patterns}). Representation patterns are derived from embedding patterns. The enumeration and nomenclature is driven by clear rules. Two major types of embedding hiding patterns exist: (1) those that modulate a state or value, e.g., the pattern \MethDescr{E1.3.\ LSB State/Value Modulation}\footnote{In this paper, we will use a highlighted font to indicate our proposed nomenclature, including the patterns introduced in \cite{CSUR25}.} subsumes LSB-based methods; (2) those that modify the occurrence of some element, e.g., the pattern \MethDescr{E2.1.\ Element Enumeration} encodes secret information through the number of some element (e.g., byte count of a file or size of a network packet) \cite{CSUR25}. A brief overview on the major patterns is given by Tab.~\ref{tab:patternsoverview}.
Note that patterns can be media-specific, e.g., \MethDescr{E1.3n1.\ Network LSB State/Value Modulation} or \MethDescr{E1.3t1.\ Text LSB State/Value Modulation} for network and text steganography, respectively.

\begin{table}[!ht]
    \centering
    \resizebox{1.00\textwidth}{!}{
    \begin{tabular}{|l|p{7cm}|}
        \hline
         \textbf{Pattern} & \textbf{Brief Description} \\ \hline  \hline
        \MethDescr{E1. State/Value Modulation} & Some state (e.g., of an actuator) or value (e.g., bit in a file) is modulated to hide a secret message. \\ \hline
        ~~\MethDescr{E1.1. Reserved/Unused State/Value Mod.} & Sub-variant where reserved or unused states/values (e.g., padding bits) are modulated to embed a secret message \\ \hline
        ~~\MethDescr{E1.2. Random State/Value Mod.} & Sub-variant that covers the modulation of random values (e.g., cryptographic hashes) \\ \hline
        ~~\MethDescr{E1.3. LSB State/Value Mod.} & Sub-variant covering all forms of LSB steganography \\ \hline
        ~~\MethDescr{E1.4. Character State/Value Mod.} & Sub-variant covering textual character modulations, e.g., changing the case of letters \\ \hline
        ~~\MethDescr{E1.5. Redundancy State/Value Mod.} & Sub-variant that covers hiding methods that change redundancy (to embed secret data), e.g., transcoding steganography \\ \hline  \hline
        \MethDescr{E2. Element Occurrence} & Some element's occurrence is changed in some way (e.g., a network packet \emph{appears}) \\ \hline
        ~~\MethDescr{E2.1. Element Enumeration} & Sub-variant covering methods where the number of elements is modulated to encode a secret message (e.g., number of bytes of a file) \\ \hline
        ~~\MethDescr{E2.2. Element Positioning} & Sub-variant covering methods where the location of an element in a cover object is used to encode a secret message (e.g., position of a specific pixel in an image or time of appearance of a signal) \\ \hline
    \end{tabular}
    }
    \caption{Overview of core hiding patterns of \cite{CSUR25}.}
    \label{tab:patternsoverview}
\end{table}

\paragraph{Local and Distributed Channels.}
Local steganography channels do not rely on distributing the secret message over multiple cover objects nor do they apply multiple hiding methods to the same cover object. The following terms were proposed in \cite{CSUR15} to describe distributed hiding methods using patterns: \emph{Pattern variation} refers to techniques that apply the same hiding pattern to different carrier objects, e.g., least significant bit modulation to a field in the IPv4 and the IPv6 header, without needed a whole new implementation. \emph{Pattern combination} applies multiple hiding patterns to the same carrier object (e.g., embedding a secret bit into the least significant bit of a timestamp in a file's metadata while embedding additional secret data into unused metadata bits). Finally, \emph{pattern hopping} refers to the (pseudo-randomized) alternation of hiding methods \cite{CSUR15}. Mazurczyk \ea slightly extended these terms in \cite{ARES2018Patterns} by further splitting pattern variation into \emph{host-}, \emph{flow-} and \emph{protocol-based scattering}, which represent approaches specific for network steganography. The core idea is to distribute secret message bits to different hosts, through multiple flows or through multiple protocols.

\paragraph{Direct and Indirect Channels.}
Some steganography methods, including network techniques for the creation of covert channels, are designed to establish indirect paths for secret messages, i.e., the covert sender does not directly send data to the covert receiver \cite{zander,zander2010performance}. For instance, personal cloud storage services can be used to implement an encoding for transferring secret information. In this case, file operations (e.g., copying and renaming) can be grouped into patterns and are used to generate suitable signaling flows to convey secret data to the intended recipients~\cite{stegodropbox}.

When the covert communication path does \emph{not} behave in a direct end-to-end flavor and is based on the (involuntary) integration of a third-party node, such as a network host or a process on a local system, the channel can be described by two indirect hiding patterns that have been introduced by Schmidbauer and Wendzel: \emph{redirector} or \emph{broker}, with the broker having the two sub-patterns \emph{proxy} and \emph{dead drop}) \cite{SchmidbauerWendzel:IndirectCCSurvey}. A brief description of these indirect hiding patterns is summarized in Tab.~\ref{tab:indirectpatterns}.

\begin{table}[!ht]
    \centering
    \resizebox{1.00\textwidth}{!}{
    \begin{tabular}{|l|p{8.1cm}|}
        \hline
         \textbf{Indirect Hiding Pattern} & \textbf{Brief Description} \\ \hline  \hline
        \MethDescr{Redirector} & A sender forces a third-party node to unintentionally redirect steganography objects to a covert receiver. Example: a covert sender transmits steganography objects as payload within a spoofed network packet. The packet contains a request (e.g., ICMP or IGMP) and is sent to a third-party node that responds to the spoofed address (which is the one of the covert receiver) \cite{SchmidbauerWendzel:IndirectCCSurvey}. \\ \hline \hline
        \MethDescr{Broker} & In comparison to the \MethDescr{Redirector}, a broker does not redirect steganography objects but manipulates the third-party node so that these steganography objects can be extracted by a covert receiver \cite{SchmidbauerWendzel:IndirectCCSurvey}. A \MethDescr{Broker} is either a \MethDescr{Proxy} or a \MethDescr{DeadDrop}. \\ \hline
        ~~\MethDescr{Proxy} & Sub-variant of the \MethDescr{Broker} where a covert sender influences a third-party node in such a way that the influence can be recognized by a covert receiver. Example: A local covert sender process might cause heavy load on a third-party process handling his requests. This load influences the third-party process' performance and can be measured by a covert receiver's process. The influence on the performance represents the secret message. \\ \hline
        ~~\MethDescr{Dead Drop} & Sub-variant of the \MethDescr{Broker} where the steganography object is stored on a third-party node. Example: a sender might influence the network protocol's cache of a third-party node to embed a secret message. The cache's content is then read by the covert receiver \cite{SchmidbauerWendzel:IndirectCCSurvey}. \\ \hline
    \end{tabular}
    }
    \caption{Summary of indirect hiding patterns as introduced in \cite{SchmidbauerWendzel:IndirectCCSurvey}; descriptions have been generalized to remove the network-specific context.}
    \label{tab:indirectpatterns}
\end{table}

\paragraph{Active and Passive Channels.}
Covert senders do not necessarily need to create their own cover objects to embed data into, nor must covert receivers necessarily be the overt recipients of a cover object \cite{ZI2010686}. In this perspective, an \emph{active} channel is one where the covert sender creates the cover object and the covert receiver is the overt recipient. In contrast, a \emph{passive} channel would require a covert sender to modify a third-party cover object to embed the secret message into and the covert receiver to recognize the embedded message (e.g., as an on-path attacker). It is also possible to create channels of mixed form (\emph{semi-active} \cite{SemiPassiveSemiActive} as introduced by Lamshöft and Dittmann;  \emph{semi-passive} as introduced by Zander \cite{zander2010performance}). In some cases \emph{fully-passive} channels can be created (as introduced by Wendzel \ea~\cite{DYST}). In such cases, the cover object is untouched by sender and receiver. Instead, the sender solely \emph{points} to the cover object and the receiver observes the traffic through eavesdropping or as a broadcast receiver.

\paragraph{Multi-level Steganography.}
Multiple authors proposed nesting steganographic objects inside other steganography objects, leading to multiple levels (or layers) of steganography. Multi-level steganography can be found in different domains, including filesystem steganography \cite{Masud_2024} and network steganography \cite{MultiLevelStego}. Multi-level steganography can be used to reach a plausible deniability, where the outer steganography level is presented to an observer, but inner levels are not revealed and kept secret.

A similar concept introduced by Ogiela and Koptyra is called \emph{multi-secret steganography} \cite{ogiela2015false}. Instead of nesting one layer inside another, multi-secret steganography embeds a set of secret messages into the same carrier. Several of these messages are \emph{false stego-objects} that are comparably easy to detect while the actual secret message is more challenging to detect.\footnote{Multi-secret steganography could alternatively be considered a special variant of the previously-mentioned \emph{pattern combination} \cite{CSUR15} as multiple hiding methods are combined to add secret message (fragments) to the same cover object.}

\paragraph{Pointers to ``Historic'' and ``Future'' Data.}
A steganography transmission can embed the actual secret message or a \emph{pointer} that refers to the desired secret message, so that only a small fraction of the information is used instead of transferring the entire message~\cite{DYST}. In the case of a pointer, one can refer to either already existing data (called historic data, even if it was \emph{just} created, e.g., a few CPU cycles ago) or to anticipated future data (e.g., expected regular ARP requests).

\paragraph{Unified Description Method.}
In 2016, a \emph{unified description method} (UDM) was introduced for network information hiding methods, which was slightly modified in 2025 to fit all domains of steganography and better integrate the updated patterns taxonomy \cite[supplemental material Sect.~A.6]{CSUR25}.
The UDM covers several attributes that are described in a comparable manner, including the application scenario, the hiding patterns of a hiding method, the required properties of a cover object that are necessary to realize the steganography channel, and the channel properties (e.g., robustness, countermeasures, and capacity). Finally, optional information can be provided on a channel-internal protocol.
The UDM has been designed to improve the replicability, comparability, and identification of research gaps in the steganography literature. 

\section{Proposal for a Combination of Description Methods}\label{sect:meta}

\begin{figure}
    \centering
    \includegraphics[width=1.0\linewidth]{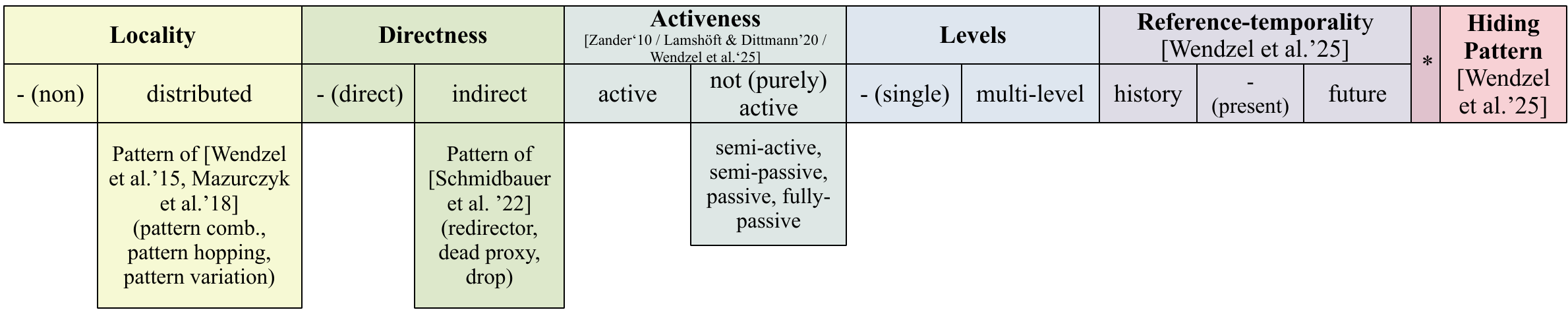}
    \caption{Proposed Naming Convention}
     \label{fig:namingconv}
\end{figure}

Fig.~\ref{fig:namingconv} shows the general structure of our naming convention. Following the recent steganography taxonomy \cite{CSUR25}, here we apply the proposed pattern naming convention for ``hiding patterns'' (see the most-right component in the figure). At the same time, we also adjust it to incorporate the surrounding terms for categorization (see the remaining boxes in the figure). A dash (-) indicates a default category, i.e., it can be omitted if a channel does not contain a specific feature. In particular, it must not be mentioned explicitly that a steganography channel is non-distributed, direct, active, uses solely a single level of embedding, or when the steganography data is embedded into the object itself (present-focused) rather than referring to history or anticipated (future) data.

All components of our naming convention are additionally explained by our inter-active online tool: \url{https://patterns.omi.uni-ulm.de/desrcovert/}

\subsection{Naming Components in Detail}

Referring to Fig.~\ref{fig:namingconv}, we now discuss the components that can be used to develop the proposed naming convention. The naming components are described below. 

\paragraph*{Locality.}
This clarifies whether a steganography hiding method is local or distributed. 
To this end, the first (but optional) component of the naming can be ``distributed'', followed by the distribution pattern (e.g., ``pattern combination'') enclosed in brackets~\cite{CSUR15,ARES2018Patterns}. Such methods might employ different hiding patterns simultaneously. As will be detailed later in Sect.~\ref{sect:example:RTTindirect}, multiple methods should be mentioned jointly.

\paragraph*{Directness.}
This clarifies if the hiding method represents a direct or indirect channel. The optional attribute ``indirect'' would be followed by the name of the particular \emph{pattern} of Schmidbauer and Wendzel \cite{SchmidbauerWendzel:IndirectCCSurvey} in brackets, e.g., ``redirector''. Multiple examples featuring such indirect patterns will be provided in Sect.~\ref{sect:examples}.

\paragraph*{Activeness.}
This defines whether the channel is \emph{active} (must not be mentioned explicitly) or \emph{passive} \cite{ZI2010686}. Here, the above-mentioned passivity terms (\emph{semi-active}, \emph{semi-passive}, \emph{fully-passive}) can be be placed in brackets behind the ``passive'' attribute.

\paragraph*{Level Characteristic.}
This denotes whether the steganography method establishes a multi-level steganography system, or not. The attribute should be omitted if it is a single-level system.

\paragraph*{Reference-temporality.}
This attribute can be used to specify if secret data is ``moved'' through the channel by means of pointers, that is if the channel points to previously found/written data (history covert channel) or to anticipated (future) data as introduced in~\cite{DYST}.

\paragraph*{Star-property (*).}
This attribute is a star property (\textbf{*}) that allows arbitrary details to be added. For instance, one might employ terms like \emph{cover selection} \cite{Fridich2009} or \emph{coverless steganography} \cite{CoverlessImageStego,CoverlessImageStego2}. Another option is to state whether a channel is a unidirectional, bidirectional, or broadcast channel. If desired, one might explicitly mention rough robustness criteria, e.g., that a channel is \emph{noisy} or \emph{noise-free} or if the cover is \emph{predictable}, \emph{variable} or \emph{randomized} as done in the work of Zander~\cite{zander2010performance}. Finally, one might state that a hiding method is \emph{reversible} \cite{chang2006reversible,HesselingKellerLitzinger23:Reversible,mazurczyk2019towards,Song:reversible2018}, i.e., if an (intermediate) covert receiver can restore the cover object to its status before a secret message was embedded.

\paragraph*{Hiding Pattern.}
This attribute must mention the hiding pattern of the 2025 taxonomy of steganography hiding methods \cite{CSUR25}. This is a very important aspect, as this attribute contributes to aligning the various concepts within the literature towards a common knowledge.

\subsection{Describing Sophisticated Methods}

In general, the aforementioned \textit{seven} attributes already provide some flexibility to describe sophisticated methods. For example, one might apply the distributed \emph{host-based scattering} method introduced in \cite{ARES2018Patterns} by using an indirect communication through \emph{dead drops} \cite{SchmidbauerWendzel:IndirectCCSurvey} that serve as nodes for storing the secret data. The actual hiding method might be a network-based LSB state-value modulation. To this end, we would call such a method a 
\MethDescr{distributed (host-based scattered) indirect (dead drop) E.1n1.\ network LSB state/value modulation}.

However, if the steganography method \emph{utilizes different hiding patterns}, each must be named separately. For instance, one might be LSB and the other reserved/unused state/value modulation, so we would gain two descriptions, (a) and (b): %
\MethDescr{distributed (host-based scattered) indirect (dead drop) (a) E1.3n1.\ network LSB state/value modulation and (b) E1.1n1.\ network reserved/unused state/value modulation)}.

Unfortunately, methods might also apply \emph{multi-level steganography}, i.e., nesting stego objects inside other stego objects. This requires more elaborate naming, but the availability of a solid convention/taxonomy makes the process simple and reduces ambiguities. In this case, each stego-layer could utilize a \emph{different} hiding pattern. For instance, a multi-layer filesystem steganography method might be classified as \MethDescr{multi-level (a) E1.3f1.\ filesystem LSB state/value modulation, (b) E1.1f1.\ filesystem reserved/unused state/value modulation and (c) E1.2f1.\ filesystem random state/value modulation)}. 

This means that the outermost layer performs LSB state/value modulation, while the middle layer performs reserved/unused state/value modulation, and the innermost layer is LSB state/value modulation on filesystem data.

It would also be reasonable to have a multi-\emph{level}-multi-\emph{media} approach. For instance, network steganography might hide data inside network packets, and embedded data could contain a second layer featuring digital image steganography data. An example of such a case would be \MethDescr{multi-level (a) E1.1n1.\ network reserved/unused state value modulation, (b) E1.3d1.\ digital media LSB state/value modulation}.

Allowing such multi-media descriptions also contributes to fill a gap identified by the latest taxonomy \cite[cf.\ Fig.~1]{CSUR25}. Especially, hiding methods have been described until now as belonging to only \emph{one} domain, i.e., neglecting another domain if they utilize objects or hiding methods from multiple domains.

\subsection{Utilization of the Unified Description Method}

\begin{figure}
    \centering
    \includegraphics[width=1.0\linewidth]{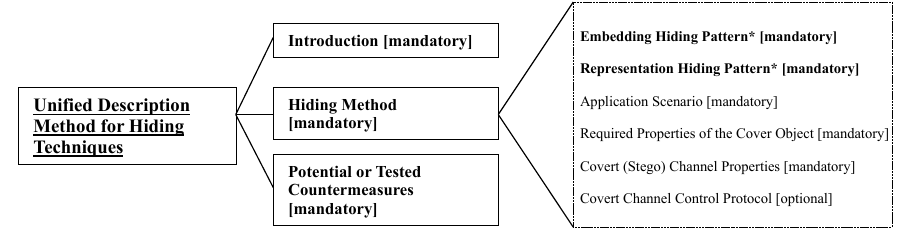}
    \caption{The UDM of \cite{CSUR25}. Our nomenclature is used within the attributes specifying the patterns (indicated by \textbf{*}).}
    \label{fig:UDM}
\end{figure}

In case our nomenclature cannot capture all the nuances of the targeted steganography approach, the missing details can be covered by borrowing ideas from the UDM \cite[cf.\ electronic supplement]{CSUR25}. Fig.~\ref{fig:UDM} shows the structure of the UDM. As shown, such a unified framework foresees that a steganography method is described by using hiding patterns as its core component as well as by several additional attributes. These additional attributes (application scenario, required properties of the cover object, etc.) leave room for a structured description of typical attributes.

\paragraph{Handling Representation Patterns.}
Some steganography methods utilize a different embedding and representation pattern. In this case, the attribute ``representation hiding pattern'' of the UDM can be used to mention the representation pattern, also following our nomenclature (see example in Sect.~\ref{sect:example:RTTindirect}). In other cases where the representation pattern is simply the representing variant of the embedding pattern, the UDM attribute ``representation hiding pattern'' can just state ``representation variant of embedding pattern''.

\section{Examples}\label{sect:examples}
This section describes existing steganography methods by using our combined approach. Tab.~\ref{tab:examples} provides an overview on our examples. We start with a simple network-specific technique, followed by two indirect/hybrid network steganography techniques. Afterwards, we cover an example from digital media steganography, one from cyber-physical systems steganography, and one from text steganography.

\begin{table}[!ht]
    \centering
    \begin{tabular}{|c|p{3.7cm}|p{7.6cm}|}
      \hline
      \textbf{Ex.} & \textbf{Scenario} & \textbf{Description} \\ \hline
       1  & Simple Network Method & \MethDescr{E1.1n1.\ Network Reserved/Unused State/Value Modulation} \\ \hline
       2  & Hybrid Hiding Method & (1) \MethDescr{Indirect (Proxy) E1n1.\ Network State/Value Modulation}, (2) \MethDescr{Indirect (Proxy) R2.2n1.\ Network Element Positioning} \\ \hline
       3 & Indirect Dead-drop & \MethDescr{Indirect (Dead Drop) E1.1n1.\ Network State/Value Modulation} \\ \hline
       4 & (History-focused) LSB Audio Steganography & \MethDescr{(History-focused) E1.3d1.\ Digital Media LSB State/Value Modulation} \\ \hline
       5 & (Distributed) OPC UA Steganography & \MethDescr{(Distributed) E1.3c1.\ CPS LSB State/Value Modulation} \\ \hline
       6 & Text Steganography & (1) \MethDescr{E2.1t1.\ Text Element Enumeration}, (2) \MethDescr{(Semi-active) E2.1t1.\ Text Element Enumeration}, (3) \MethDescr{Indirect (Dead-drop) E2.1t1.\ Text Element Enumeration} \\ \hline
    \end{tabular}
    \caption{Overview on provided examples}
    \label{tab:examples}
\end{table}

\subsection{Simple Network Steganography Method}

In network steganography, secret data is hidden inside the content of network packets (e.g., overwriting unused bits or replacing values in header fields) or by modulating temporal characteristics of the traffic (e.g., influencing packet occurrence or flow duration) \cite{mileva2014covert,CSUR15,zander}. The first methods for network steganography (or: network covert channels) have been described in the late 1980s, e.g., \cite{Girling87,WolfCCsLAN}. Recently, network steganography has become a major branch of steganography-capable malware \cite{Strachanski:Stegomalware,caviglione2022never}.

In this first example, we assume that a covert sender hides data within the unused ``reserved'' bit of the IPv4 header. Without any additional sophisticated characteristics, such as multi-level steganography or an indirect manipulation of values, the whole description consists of the pattern \MethDescr{E1.1n1.\ Network Reserved/Unused State/Value Modulation}. As the covert receiver interprets the same reserved bit of the IPv4 header, the representation pattern is \emph{R1.1n}.

\subsection{Hybrid Hiding Method}\label{sect:example:RTTindirect}

Spiekermann \ea \cite{spiekermann2017towards} present a migration covert channel. They propose that a covert sender migrates a virtual machine from one server to another, e.g., from Europe to Australia, so that a covert receiver can measure a different round trip time (RTT). While the embedding hiding pattern to migrate the virtual machine is a \MethDescr{E1n1.\ Network State/Value Modulation} (commands in a protocol are transferred so that the migration is triggered), the covert receiver must measure the RTT with probe traffic or by conducting some measurements. In particular, the time of occurrence of a response packet is measured to obtain the RTT, which is \MethDescr{R2.2n1.\ Network Element Positioning} (the network element (=packet) is ``positioned'' in time \cite{CSUR25}).

To map the underlying patterns of this ``migration'' channel into the proposed nomenclature, we utilze Fig.~\ref{fig:namingconv} as a guide. The hiding method is \emph{not} distributed, but it is an indirect method following the ``proxy'' pattern \cite{SchmidbauerWendzel:IndirectCCSurvey}. The method is an active method without multi-level component, and no history/future reference, i.e., these three aspects must not be mentioned explicitly. Thus, we call this hybrid method as follows: \MethDescr{Indirect (Proxy) E1n1.\ Network State/Value Modulation} (the embedding pattern) and \MethDescr{Indirect (Proxy) R2.2n1.\ Network Element Positioning} (the representation pattern).
Note that the indirect attribute is included in both, embedding and representation patterns.

\subsection{Network-based Indirect Dead-Drop Hiding Method}
Velinov \ea introduced an indirect method that allows the establishment of a bidirectional network covert channel in the MQTT protocol, which has been described as ``ICC.1'' in \cite{DBLP:journals/access/VelinovMWM19}. The channel is also referred to as ``MQTT.1'' in \cite{SchmidbauerWendzel:IndirectCCSurvey}. MQTT is specialized in conveying information between IoT devices, the clients of the MQTT server, via topics. Such topics may have numerous subtopics and can be subscribed to by clients. To achieve indirect covert communication, the covert sender and covert receiver (i.e., clients of the MQTT server) must agree on one first-level topic. The covert receiver has to subscribe to all subtopics of this first-level topic. The covert sender now embeds covert information in one of the subtopics that is stored by the MQTT-server. Due to subscription, the MQTT-server notifies the covert receiver and sends the embedded covert information, which can be extracted at the target of the covert channel. Due to the storing of information on the MQTT server, \cite{SchmidbauerWendzel:IndirectCCSurvey} considers this concept to be a so-called dead drop. For such an implementation, the covert channel is split into three phases: manipulation, storing, and extraction. Each phase may utilize different hiding methods, but in the case of this example, each phase uses the patterns E1n1.\ and R1n1.\ for embedding and representation of covert information, respectively. 

In the proposed nomenclature, this covert channel can be defined as \emph{non-}distributed but ``indirect'' (using the ``dead drop'' pattern). Furthermore, the method is active, does not apply multi-level steganography, and does not employ a history/future component. Thus, the complete categorization is \MethDescr{Indirect (Dead Drop) E1.1n1.\ Network State/Value Modulation}.

\subsection{Audio Hiding Method}
Despite dating back to the late 1990s, Least Significant Bit (LSB) modification is still one of the most relevant hiding methods used in audio steganography. From the approximately 800 audio steganography repositories currently available on Github, an estimated 70\% features LSB modification based hiding methods, such as LSB Replacement (LSBR) and also one of the few audio stego-malware cases reported in the wild
features LSBR as the embedding method~\cite{caviglione2022never,Strachanski:Stegomalware}.

Technically, in LSBR parts of the least significant bit plane of an uncompressed audio or image sample is adapted to match the bits in a secret message. In compact disc (CD) compliant pulse code modulated (PCM) WAV audio files, every sample value is encoded by a 16 bit value per channel. In that case, the lowest bit-plane value is modified, producing minimal changes in the audio content that are much below the hearing threshold of a human being, even if introduced into silent parts of the audio content. In general, LSBR for image and audio is considered to be easily detectable if high embedding rates (>20\% of potential hiding places) are used since the message embedding in those cases overwrites/destroys the underlying characteristics of the cover signal. For very low embedding rates (\mbox{i.e.,} very short messages hidden in long covers) the introduced embedding artifacts have been shown to be statistically indistinguishable from the cover statistics and therefore undetectable.

Due to the prominence of this method in digital media steganography the pattern \MethDescr{E1.3d1.\ Digital Media LSB State/Value Modulation} has been proposed \cite{CSUR25}. A hiding method would simply be described by its pattern if no sophisticated approach is given.

Instead, if a hiding method would only embed pointers to previously recorded data in a cover object, it would be categorized as \MethDescr{History-focused E1.3d1.\ Digital Media LSB State/Value Modulation}.

\subsection{OPC UA Hiding Method}

Neubert \ea present and analyze three methods for steganographic embedding of hidden messages in OPC UA data packets \cite{neubert2024}. OPC UA is a cross-platform, open-source protocol for Cyber-Physical Systems (CPS), in particular for Industrial Control System (ICS) network communication, and is developed by the OPC Foundation \cite{OPCFoundation}. The goal of Neubert \ea is to generate and evaluate steganographic  OPC UA network traffic including packets generated by simulation of a corrupted Programmable Logic Controller (PLC) within an ICS network.
The PLC generates OPC UA packets with slight time\-stamp modifications in micro- and nanosecond range to embed hidden messages. OPC UA timestamps are composed in the format \textit{"T\textsubscript{i} = hh:mm:ss:mmm µµµ nnn”}, where h,m,s,m,µ and n stand for hour, minute, second, millisecond, microsecond and nanosecond respectively. 
All three methods make use of the least two digits of the microsecond timestamp and all three digits of the nanosecond values. This results in timestamps such as \textit{"T\textsubscript{i} = 10:00:00.123 4\textbf{56 789}”} for each modified packet, where the five potentially modified digits are marked in bold, \textit{“56”} representing the lower 2 digits of the microseconds and \textit{“789”} the three digits for the nanoseconds. The three algorithms vary with respect to their strategy for constructing the actual embedding pattern and its position. The first and simplest method generates patterns by embedding two distinct digit values as embedding symbols for 1 and 0 for each hidden message bit in all three positions for the microsecond timestamp of three subsequent OPC UA packets (i.e. positions of \textit{“456”} in the above example) in the communication flow. A second method involves basically the same scheme, but performs an embedding key-based permutation of the embedding symbols, as well as the embedding digit positions. The third method additionally involves the timestamp values at positions not considered for modification to generate patterns and uses XOR encryption.

All three examples fall in the category \MethDescr{E1.3c1.\ CPS LSB State/Value Modulation} because the least significant digits (although not exactly bits!) of timestamp elements are modulated. 
All other properties can be omitted, as they fall into the default categories, because the payload is bit-wise directly represented, all channels are active (due to the assumption of a compromised PLC component), do not consider multi-level steganography and carry the message directly. Thus, the directness, activeness, level characteristic and reference-temporality properties can be omitted. As a result, the simplest method can be described as \MethDescr{Non-Distributed E1.3c1.\ CPS LSB State/Value Modulation}, whereas the remaining two methods fall into the category of \MethDescr{Distributed E1.3c1.\ CPS LSB State/Value Modulation}.
The two distributed methods utilize a key-based permutation of the embedding position of each single numeric symbol within the 5 potential least significant numeric positions (i.e., \textit{“56789”} in the above example) across subsequent data packets.
The description for the pattern combination of the two distributed methods can be formulated as \textit{“key-based symbol, embedding position and cover data permutation“} using the *-property of Fig.~\ref{fig:namingconv}.

\subsection{Text Hiding Method}
Most of the methods applied in text steganography are simple. For example, repeating white space characters in a text (open space method) \cite{Bender:1996,Ahvanooey:2019} is a form of the pattern \MethDescr{E2.1t1.\ Text Element Enumeration}. However, these methods can be applied in heterogeneous scenarios, with different naming components, such as directness and activeness. For example, Mileva \ea \cite{Mileva2021DICOM} suggest three different applications of the open space method and the pattern E2.1t1 using DICOM files as covert carriers. DICOM (\emph{Digital Imaging and Communications in Medicine}, cf.\ \cite{DICOMstd}) is a standard for the digital handling of medical images, containing several attributes that can be filled with textual content. 

In the first scenario, a covert sender and a covert receiver are the actual sender and receiver, which means that the created covert channel is active and direct, so we can describe it simply as \MethDescr{E2.1t1.\ Text Element Enumeration}. The second scenario covers a direct channel in which the covert sender is the actual sender, while covert receiver(s) monitor the network traffic intended for other receivers to extract the hidden message. This covert channel can be described as \MethDescr{Semi-active E2.1t1. Text Element Enumeration}. The last scenario is an example of an indirect active channel in which the covert sender (as the actual sender) stores the cover DICOM file in an archive (utilized as third-party intermediate component), while the covert receiver(s) can request some services regarding that DICOM file from the archive and extract the hidden message. Thus, this hidden communication mechanism can be described as \MethDescr{Indirect (Dead-drop) E2.1t1.\ Text Element Enumeration}.

\section{Discussion}\label{sect:disc}
The examples provided here do not cover the full extent of existing hiding methods, such as steganography in filesystems \cite{HanPGP10,Eckstein:2005}, AI models \cite{cansAI,SongAI,stegonet}, air-gapped covert channels \cite{Guri:BitWhisper,carrara2016out} or covert channel-based traffic obfuscation for censorship circumvention \cite{Xie:2024:CCinDomainFronting}. However, we believe that our methodology can be easily extended to these domains, especially since the patterns taxonomy \cite{CSUR25} already foresees many of them. At the same time, we are also aware that our work might be limited due to the lack of some categorizations and subtaxonomies of information hiding topics. However, the proposed approach and the related corpus of works at the basis of \cite{CSUR25} are solid, thus making improvements and adjustments easier. As our work becomes accepted by the community, routine ``maintenance'' operations will be easier. In this perspective, information hiding topics that may emerge in the future due to the utilization of new technologies not yet invented or not yet relevant could be added when needed. As an example, the vivid area of tracking AI-generated content was completely unforeseeable ten years ago but now drives vast research directions. This reinforces the need of having multiple and flexible attributes to describe how the steganography area evolves. 

Another limitation of our work is that it does not cover countermeasures; they are solely an attribute of the UDM. However, since our work is focused on the categorization and description of hiding methods instead of the categorization and description of countermeasures to detect, limit, or prevent steganography, we consider the description of countermeasures as a separate project. 

Future developments are notoriously difficult to predict. For this reason, our approach might be considered an intermediate step that allows extension in areas where it is needed (due to its *-property in Fig.~\ref{fig:namingconv}).

\section{Conclusion}\label{sect:concl}
We have introduced a meta-view on description and taxonomy approaches in steganography that several researchers focusing on information hiding have constructed over the years. Our combination of existing methodologies allows for a comprehensive description of steganography methods in a unified and comparable fashion. As a result, scientific re-inventions could be reduced and the knowledge between different fields (e.g., defensive watermarking and detection of stegomalware) could be exchanged much more effectively. Our provided inter-active online tool aids the understanding of our methodology. Additionally, the tool can be used for didactic settings.

Future work will focus on the development of a similar methodology for countermeasures against threats endowed with some form of steganographic capabilities. Specifically, we are working towards the definition of a suitable abstraction/taxonomy to prevent ambiguities and imperfect isolation issues that may lead to exploitable hiding patterns. 

\begin{credits}
\subsubsection{\ackname} 
We like to thank the anonymous reviewers for their constructive feedback. We further like to express our gratitude to co-authors of previous works that helped paving the way to this paper: Wojciech Mazurczyk, Jörg Keller, Krzysztof Cabaj, Laura Hartmann, Sebastian Zillien, Tom Neubert, Bernhard Fechner, and Christian Herdin.

The work of Luca Caviglione has been been supported by PNRR project.

\subsubsection{\discintname}
The authors have no competing interests to declare that are relevant to the content of this article.
\end{credits}

\bibliographystyle{splncs04}
\bibliography{papers}

\end{document}